\documentclass[12pt]{article}
\usepackage[english,german,french,polish]{babel}
\usepackage[T1]{fontenc}

\begin{document}

\baselineskip 0.75cm
\topmargin -0.6in
\oddsidemargin -0.1in

\let\ni=\noindent

\renewcommand{\thefootnote}{\fnsymbol{footnote}}

\newcommand{\CKM}{Cabibbo--Kobayashi--Maskawa }

\newcommand{\SM}{Standard Model }

\pagestyle {plain}

\setcounter{page}{1}

\pagestyle{empty}

~~~

\begin{flushright}
IFT--01/07
\end{flushright}

{\large\centerline{\bf Toward neutrino texture dominated by}}
{\large\centerline{\bf Majorana lefthanded mass matrix{\footnote {Supported in part by the Polish KBN--Grant 5 P03B 119 20 (2001--2002).}}}}

\vspace{0.3cm}

{\centerline {\sc Wojciech Kr\'{o}likowski}}

\vspace{0.2cm}

{\centerline {\it Institute of Theoretical Physics, Warsaw University}}

{\centerline {\it Ho\.{z}a 69,~~PL--00--681 Warszawa, ~Poland}}

\vspace{0.3cm}

{\centerline{\bf Abstract}}

\vspace{0.2cm}

A form of mixing matrix for three active and three sterile, conventional Majorana neutrinos is proposed. Its Majorana lefthanded part arises from the popular bimaximal mixing matrix for three active neutrinos that works satisfactorily in solar and atmospheric experiments if the LSND effect is ignored. One of three sterile neutrinos, effective in the Majorana righthanded and Dirac parts of the proposed mixing matrix, is responsible perturbatively for the possible LSND effect by inducing one of three extra neutrino mass states to exist actively. The corresponding form of neutrino mass matrix is derived. If all three extra neutrino mass states get vanishing masses, the neutrino mass matrix is dominated by its specific Majorana lefthanded part. Then, the observed qualitative difference between mixings of neutrinos and down quarks may be connected with this Majorana lefthanded dominance realized for neutrinos.

\vspace{0.2cm}

\ni PACS numbers: 12.15.Ff , 14.60.Pq , 12.15.Hh .

\vspace{0.6cm}

\ni March 2001

\vfill\eject

~~~
\pagestyle {plain}

\setcounter{page}{1}

\vspace{0.2cm}

\ni {\bf 1. Introduction}

\vspace{0.2cm}

Although the recent experimental results for atmospheric $\nu_\mu $'s as well as solar $ \nu_e $'s are in favour of excluding the hypothetical sterile neutrinos from neutrino oscillations [1], the problem of the third neutrino mass difference manifested in the possible LSND effect for accelerator $\nu_\mu $'s  still exists [2], implying a further discussion on mixing of sterile neutrinos with three active flavors $ \nu_e\,, \,\nu_\mu\,, \,\nu_\tau $. In the present note we contribute to this discussion by constructing a particular $6\times 6$ texture involving three active and three sterile, conventional Majorana neutrinos. The construction extends (or rather adapts) the familiar bimaximal $3\times 3$ texture [3] working in a satisfactory way for three active neutrinos in solar and atmospheric experiments if the LSND effect is ignored. Then, one of three sterile neutrinos is responsible perturbatively [4] for the possible LSND effect by inducing one of three extra neutrino mass states to exist actively.
 
As is well known, three sterile Majorana neutrinos

\begin{equation} 
\nu^{(s)}_\alpha  = \nu_{\alpha R} + \left( \nu_{\alpha R} \right)^c \;\; (\alpha = e\,,\,\mu\,,\,\tau) 
\end{equation} 

\ni can be always constructed in addition to three active Majorana neutrinos 
 
\begin{equation} 
\nu^{(a)}_\alpha  = \nu_{\alpha L} + \left( \nu_{\alpha L} \right)^c \;\; (\alpha = e\,,\,\mu\,,\,\tau) 
\end{equation} 

\ni if there are righthanded neutrino states $\nu_{\alpha R} $ beside their familiar lefthanded partners $\nu_{\alpha L}$ participating in \SM gauge interactions [of course, $\nu^{(a)}_{\alpha L} = \nu_{\alpha L}$ and $\nu^{(s)}_{\alpha L} = (\nu_{\alpha R})^c $]. Whether such sterile neutrino states are physically realized depends on the actual shape of the neutrino mass term whose generic form is

\begin{eqnarray}
-{\cal L}_{\rm mass} & = & \frac{1}{2}\sum_{\alpha \beta} (\overline{\nu_\alpha^{(a)}} \,,\, \overline{\nu_{\alpha}^{(s)}}) \left( \begin{array}{cc} M^{(L)}_{\alpha \beta} & M^{(D)}_{\alpha \beta} \\ M^{(D)*}_{\beta \alpha}  & M^{(R)}_{\alpha \beta} \end{array} \right) \left( \begin{array}{c} \nu^{(a)}_\beta \\ \nu^{(s)}_{\beta} \end{array} \right) \nonumber \\ & = & \frac{1}{2}\sum_{\alpha \beta} \left( M^{(D)}_{\alpha \beta} + M^{(D)*}_{\alpha \beta}\right) \left( \overline{\nu_{\alpha L}}\; \nu_{\beta R} + \overline{\nu_{\beta R}}\; \nu_{\alpha L}\right)  \nonumber \\ & & + \frac{1}{2} \sum_{\alpha \beta} M^{(L)}_{\alpha \beta} \left[  \overline{\nu_{\alpha L}}\,( {\nu}_{\beta L} )^c + \overline{(\nu_{\alpha L})^c}\; \nu_{\beta L}\right] \nonumber \\ & & + \frac{1}{2} \sum_{\alpha \beta} M^{(R)}_{\alpha \beta} \left[ \overline{\nu_{\alpha R}}\, ( {\nu}_{\beta R} )^c + \overline{(\nu_{\alpha R})^c}\; \nu_{\beta R}\right]\; ,
\end{eqnarray} 

\ni where$M_{\beta \alpha}^{(L,R)*} = M_{\alpha \beta}^{(L,R)}$, but $M_{\beta \alpha}^{(D)*} \neq M_{\alpha \beta}^{(D)}$ in general. The $6\times 6$ neutrino mass matrix 

\begin{equation}
M = \left( \begin{array}{cc} M^{(L)} & M^{(D)} \\ M^{(D)\dagger}  & M^{(R)} \end{array} \right) 
\end{equation}

\ni appearing in Eq. (3) is hermitian, $ M^\dagger = M $. Here, $  M^{(D, L, R)} =\left(M^{(D, L, R)}_{\alpha \beta}\right) $ are $3\times 3 $ neutrino mass matrices: Dirac, Majorana lefthanded and Majorana righthanded, respectively. Further on, for six neutrino flavor states we will use the notation $\nu_\alpha \equiv \nu_{\alpha}^{(a)}$ and $\nu_{\alpha_s} \equiv \nu_{\alpha}^{(s)}$ with $ \alpha = e\,,\,\mu\,,\,\tau $ and then pass to $ \nu_\alpha = \nu_e\,, \,\nu_\mu\,, \,\nu_\tau \,,\, \nu_{e_s}\,, \,\nu_{\mu_s}\,, \,\nu_{\tau_s}$ where $ \alpha = e\,,\,\mu\,,\,\tau \,,\, e_s\,,\,\mu_s\,,\,\tau_s $. Six neutrino mass states will be denoted as $ \nu_i = \nu_1 \,,\,\nu_2 ,\,\nu_3 \,,\, \nu_4 \,,\,\nu_5 \,,\,\nu_6 $  where $ i = 1 \,,\, 2 \,,\, 3 \,,\, 4 \,,\,5 \,,\,6 $.  

\vspace{0.2cm}

\ni {\bf 2. Proposal of a $6\times 6$ neutrino mixing matrix}

\vspace{0.2cm}

Starting from the phenomenologically useful bimaximal mixing matrix for three active neutrinos [3,5]

\begin{equation} 
U^{(3)} =   \left( \begin{array}{ccc} \frac{1}{\sqrt{2}} & \frac{1}{\sqrt{2}} & 0 \\ -\frac{1}{2} &  \frac{1}{2} & \frac{1}{\sqrt{2}} \\ \frac{1}{2} & -\frac{1}{2} & \frac{1}{\sqrt{2}}  \end{array} \right) \;,
\end{equation} 

\ni we propose the following form of the $6\times 6$ neutrino mixing matrix:

\begin{equation} 
U =  \left( \begin{array}{cc} U^{(3)} & 0 \\ 0 & 1^{(3)} \end{array} \right)  \left( \begin{array}{cc} C & S \\ -S & C \end{array} \right) =  \left( \begin{array}{cc} U^{(3)}C & U^{(3)}S \\ -S & C \end{array} \right)   \;,
\end{equation} 

\ni where 

\begin{equation} 
C =  \left( \begin{array}{ccc} c_1 & 0 & 0 \\ 0 & c_2 & 0 \\ 0 & 0 & c_3 \end{array} \right) \;,\;  S =  \left( \begin{array}{ccc} s_1 & 0 & 0 \\ 0 & s_2 & 0 \\ 0 & 0 & s_3 \end{array} \right) \;,\;  1^{(3)} =  \left( \begin{array}{ccc} 1 & 0 & 0 \\ 0 & 1 & 0 \\ 0 & 0 & 1 \end{array} \right) 
\end{equation} 

\ni with $ c_i = \cos \theta_i \geq 0 $ and $ s_i = \sin \theta_i \geq 0\;\,(i = 1,2,3)$. One may also denote $s_1 \equiv s_{14}\,,\,s_2 \equiv s_{25}\,,\,s_3 \equiv s_{36}$, while $ s_{12} = 1/\sqrt{2}\,,\,s_{23} = 1/\sqrt{2}\,,\,s_{13} = 0$. Explicitly,

\begin{equation} 
U =\left( U_{\alpha i} \right) =  \left( \begin{array}{cccccc} \frac{c_1}{\sqrt{2}} & \frac{c_2}{\sqrt{2}} & 0 & \frac{s_1}{\sqrt{2}} & \frac{s_2}{\sqrt{2}} & 0  \\ -\frac{c_1}{2} & \frac{c_2}{2} & \frac{c_3}{\sqrt{2}} & -\frac{s_1}{2} & \frac{s_2}{2} & \frac{s_3}{\sqrt{2}} \\ \frac{c_1}{2} & -\frac{c_2}{2} & \frac{c_3}{\sqrt{2}} & \frac{s_1}{2} & -\frac{s_2}{2} & \frac{s_3}{\sqrt{2}}\\ -s_1 & 0 & 0 & c_1 & 0 & 0 \\ 0 & -s_2 & 0 & 0 & c_2 & 0 \\ 0 & 0 & -s_3 & 0 & 0 & c_3 \end{array} \right) \;,
\end{equation}

\ni where $ \alpha = e\,,\,\mu\,,\,\tau \,,\, e_s\,,\,\mu_s\,,\,\tau_s $ and  $ i = 1 \,,\, 2 ,\, 3 \,,\, 4 \,,\,5 \,,\,6 $. The relation

\begin{equation} 
\nu_\alpha  = \sum_i U_{\alpha i}  \nu_i 
\end{equation}

\ni describes the mixing of six neutrinos.

In the representation where the mass matrix of three charged leptons $ e^-\,,\, \mu^-\,,\,\tau^-$ is diagonal, the $6\times 6$ neutrino mixing matrix $ U $ is at the same time the diagonalizing matrix for the $6\times 6$ neutrino mass matrix $ M = \left( M_{\alpha \beta} \right)$:

\begin{equation} 
U^\dagger M U = {\rm diag}(m_1\,,\,m_2\,,\,m_3\,,\,m_4\,,\,m_5\,,\,m_6)\;,
\end{equation} 

\ni where we put $m^2_1 \leq m^2_2 \leq m^2_3 $ and {\it either} $m^2_3 \leq m^2_4  $ {\it or} $m^2_ 4 \leq m^2_1  $. Then, evidently, $ M_{\alpha \beta} = \sum_i U_{\alpha i} m_i U^*_{i \beta} $. From this formula, we obtain with the use of proposal (8) the following particular form of $ 6 \times 6$ neutrino mass matrix (4):

\begin{eqnarray} 
\!\!\!\!M & \!\!\!=\!\! &\!\! \left( \begin{array}{cc} M^{(L)} & M^{(D)} \\ M^{(D)\dagger}  & M^{(R)} \end{array} \right) = \left( M_{\alpha \beta} \right) \nonumber \\
& \!\!\!=\!\! &\!\! \left( \begin{array}{cccccc} M_{e e} & M_{e \mu}& -M_{e \mu}\;\; & M_{e e_s}& M_{e \mu_s} & 0  \\ M_{e \mu} & M_{e e} + M_{\mu \tau}& M_{\mu \tau} & -M_{e e_s}/\sqrt{2}\;\; & M_{e \mu_s}/\sqrt{2} & M_{\mu \tau_s} \\ -M_{e \mu}\;\; & M_{\mu \tau} & M_{e e}+M_{\mu \tau} & M_{e e_s}/\sqrt{2} & -M_{e \mu_s} /\sqrt{2}\;\; & M_{\mu \tau_s} \\ M_{e e_s } & -M_{e e_s}/\sqrt{2}\;\; & M_{e e_s} /\sqrt{2} & M_{e_s e_s} & 0 & 0 \\ M_{e \mu_s} & M_{e \mu_s} /\sqrt{2} & -M_{e \mu_s}/\sqrt{2}\;\; & 0 & M_{\mu_s \mu_s} & 0 \\ 0 & M_{\mu \tau_s}& M_{\mu \tau_s} & 0  & 0 & M_{\tau_s \tau_s} \end{array} \right) ,\;
\end{eqnarray} 

\ni where

\vspace{-0.2cm}

\begin{eqnarray} 
M_{e e} & = & \frac{1}{2}\left(\! c^2_1 m_1 + c^2_2 m_2 + s^2_1 m_4 + s^2_2 m_5 \right)\;, \nonumber \\ M_{\mu \mu} \!=\! M_{\tau \tau}\! =\! M_{e e} + M_{\mu \tau} & =
& \frac{1}{4} \left(\! c^2_1 m_1 + c^2_2 m_2 + 2c^2_3 m_3 + s^2_1 m_4 + s^2_2 m_5 + 2 s^2_3 m_6 \right)\;, \nonumber \\ M_{e \mu}  = - M_{e \tau} & = & \frac{1}{2\sqrt{2}}\left( -c^2_1 m_1 + c^2_2 m_2 - s^2_1 m_4 + s^2_2 m_5 \right)\;, \nonumber \\  M_{\mu \tau} & = & \frac{1}{4}\left(\! -c^2_1 m_1\! - c^2_2 m_2 + 2c^2_3 m_3\! - s^2_1 m_4\! - s^2_2 m_5 + 2 s^2_3 m_6 \right) 
\end{eqnarray} 

\ni and

\vspace{-0.2cm}

\begin{eqnarray}
\frac{1}{\sqrt{2}}M_{e e_s}= - M_{\mu e_s} = M_{\tau e_s} = \frac{c_1 s_1}{2}\left( -m_1+ m_4 \right) & , & M_{e_s e_s} = s^2_1m_1+c^2_1 m_4 \;\;, \nonumber \\ \frac{1}{\sqrt{2}}M_{e \mu_s}= M_{\mu \mu_s} = - M_{\tau \mu_s} = \frac{c_2 s_2}{2}\left( -m_2+ m_5 \right) & , & M_{\mu_s \mu_s}\! =  s^2_2m_2+c^2_2 m_5 \;\;, \nonumber \\ M_{\mu \tau_s} = M_{\tau \tau_s} = \frac{c_3 s_3}{\sqrt{2}}\left( -m_3+ m_6 \right) & , &  M_{\tau_s \tau_s} = s^2_3m_3+c^2_3 m_6 \;\; ,
\end{eqnarray}

\ni while

\begin{equation} 
M_{e \tau_s} = 0 \;,\;\; M_{e_s \mu_s} = M_{e_s \tau_s} = M_{\mu_s \tau_s} = 0
\end{equation}

\ni (of course, $ M_{\alpha \beta} = M_{\beta \alpha}$ for all $\alpha $ and $\beta $). Hence, 

\begin{eqnarray}
M_{e e} - M_{e \mu}\sqrt{2} \pm M_{e_s e_s} & = & \left\{ \begin{array}{l} m_1 + m_4 \\ (c^2_1 - s^2_1)(m_1 - m_4) \end{array}\;, \right. \nonumber \\
M_{e e} + M_{e \mu}\sqrt{2} \pm M_{\mu_s \mu_s} & = & \left\{ \begin{array}{l} m_2 + m_5 \\ (c^2_2 - s^2_2)(m_2 - m_5) \end{array}\;, \right. \nonumber \\ M_{\mu \mu} + M_{\mu \tau} \pm M_{\tau_s \tau_s} & = & \left\{ \begin{array}{l} m_3 + m_6 \\ (c^2_3 - s^2_3)(m_3 - m_6) \end{array}\;, \right.
\end{eqnarray}

\ni where $M_{\mu \mu} = M_{e e} + M_{\mu \tau}$. After a simple calculation we get from Eqs. (15) 

\begin{equation} 
m_{1,4} = \frac{M_{e e} - M_{e \mu}\sqrt{2} + M_{e_s e_s}}{2} \pm \sqrt{ \left( \frac{M_{e e} - M_{e \mu}\sqrt{2} - M_{e_s e_s}}{2} \right)^{\!2} + 2 M^2_{e e_s} } 
\end{equation}

\ni and analogical formulae for $m_{2,5}$ and $m_{3,6}$ (note that $ m_1 > m_4 $, but not always $m_4 > 0 $, and similarly for $m_{2,5}$ and $m_{3,6}$). 

In the $6\times 6 $ matrix (11) there are generally nine independent nonzero matrix elements. If $s_2 = 0 $ and $s_3 = 0 $ (what implies complete decoupling of two sterile neutrinos $\nu_{\mu_s}$ and $\nu_{\tau_s}$), this number is reduced to seven. In this case, Eqs. (13) and (15) give 

\vspace{-0.3cm}

\begin{equation} 
M_{e \mu_s} = M_{\mu \mu_s} = M_{\tau \mu_s} = 0 \;\;,\;\; M_{\mu \tau_s} = M_{\tau \tau_s} = 0 \;\;,\;\; M_{\mu_s \mu_s} = m_5 \;\;,\;\; M_{\tau_s \tau_s} = m_6 
\end{equation}

\ni and 

\vspace{-0.3cm}

\begin{equation} 
M_{e e} + M_{e \mu}\sqrt{2} = m_2\;\;,\;\; M_{\mu \mu} + M_{\mu \tau} = m_3 \;\;,
\end{equation}

\ni but the formulae (16) for $ m_1 $ and $m_4 $ are not much simplified (unless $ M_{e e_s} = 0 $ {\it i.e.}, $c_1 s_1 = 0 $). Then, from Eq. (11)

\begin{equation} 
M^{(D)} =  \left( \begin{array}{rcc} -\frac{c_1 s_1}{\sqrt{2}}(m_1 - m_4) & 0 & 0 \\ \frac{c_1 s_1}{2}(m_1 - m_4) & 0 & 0 \\ -\frac{c_1 s_1}{2}(m_1 - m_4) & 0 & 0  \end{array} \right) \;\;,\;\;  M^{(R)} =  \left( \begin{array}{ccc} s^2_1 m_1 + c^2_1 m_4 & 0 & 0 \\ 0 & m_5 & 0 \\ 0 & 0 & m_6 \end{array} \right) 
\end{equation} 

\ni and $  M^{(L)}_{ee} = M_{ee} = \frac{1}{2}( c^2_1 m_1 + s^2_1 m_4 + m_2) $,  etc. If $ c_1 > s_1 \sqrt{2}$ and $ s^2_1 m_1 \gg c^2_1 |m_4| $ ({\it i.e.}, $m_1 \gg |m_4|$) and, in addition, $m_5$ and $m_6$ are vanishing, the texture is in a way of a type opposite to the see--saw (now, symbolically $(L) > (D) > (R)$ or even  $(L) \gg (D) \gg (R)$ if $ c^2_1 \gg s^2_1$ and $ m_4 = 0 $). 

At any rate, the active existence of extra massive neutrino $\nu_4 $ (in addition to the massive $\nu_1\,,\,\nu_2 \,,\, \nu_3$) is induced by the sterile neutrino $\nu_{e_s}$ mixing with the active $ \nu_e \,,\, \nu_\mu \,,\, \nu_\tau $. Of course, two completely decoupled sterile neutrinos $\nu_{\mu_s}$ and $\nu_{\tau_s}$ (with $s_2 = 0$ and $s_3 = 0$) induce trivially the passive existence of two massive neutrinos $\nu_5 = \nu_{\mu_s}$ and $ \nu_6 = \nu_{\tau_s}$ with masses $m_5 $ nad $m_6$ which, most naturally, ought to be put zero. However, another point of view is not excluded that there is still a tiny mixing of $\nu_{\mu_s}$ and $\nu_{\tau_s}$ with the rest of six neutrino flavors, caused by spontaneously breaking a GUT symmetry at a high mass scale and so, accompanied by large masses $|m_5| $ and $|m_6|$. If instead of $|m_4| \ll m_1$ there is $m_1 \ll |m_4| $, such an inequality may be not so impressive as in the familiar see-saw referring to the GUT mass scale: it may happen, for instance, that $|m_4| \sim 1$ eV and $m_1 \sim 10^{-4}$ eV; {\it cf.} Eq. (33) as an alternative to the more natural Eq. (35).
 
\vspace{0.2cm}

\ni {\bf 3. Six--neutrino oscillations}

\vspace{0.2cm}

Due to mixing of six neutrino fields described by Eq. (9), neutrino states mix according to the relation

\vspace{-0.2cm}

\begin{equation} 
|\nu_\alpha \rangle = \sum_i U^*_{\alpha i} |\nu_i \rangle \;.
\end{equation}

\ni This implies the following familiar formulae for probabilities of neutrino oscillations $ \nu_\alpha \rightarrow \nu_\beta $ on the energy shell:

\begin{equation} 
P(\nu_\alpha \rightarrow \nu_\beta) = |\langle\beta| e^{i PL} |\alpha \rangle |^2 = \delta _{\beta \alpha} - 4\sum_{j>i} U^*_{\beta j} U_{\beta i} U_{\alpha j} U^*_{\alpha i} \sin^2 x_{ji} \;,
\end{equation}

\ni being valid if the quartic product $ U^*_{\beta j} U_{\beta i} U_{\alpha j} U^*_{\alpha i}$ is real, what is certainly true when the tiny CP violation is ignored. Here,

\vspace{-0.1cm}

\begin{equation} 
x_{ji} = 1.27 \frac{\Delta m^2_{ji} L}{E} \;\;,\;\;  \Delta m^2_{ji} = m^2_j - m^2_i
\end{equation}

\ni with $\Delta m^2_{ji}$, $L$ and $E$ measured in eV$^2$, km and GeV, respectively ($L$ and $E$ denote the experimental baseline and neutrino energy, while $ p_i = \sqrt{E^2 - m_i^2} \simeq E -m^2_i/2E $ are eigenvalues of the neutrino momentum $P$).

With the use of proposal (8) for the $6\times 6$ neutrino mixing matrix and under the assumption that $s_2 = 0$ and $s_3= 0$ the oscillation formulae (21) give

\vspace{-0.2cm}

\begin{eqnarray}
P(\nu_e\, \rightarrow \nu_e)\! & \!\!=\!\! & \! 1 \!\!-\!  c^2_1 \sin^2 \!x _{21} \!-\! (c_1 s_1)^2 \sin^2 \!x _{41} \!\!-\! s^2_1 \sin^2\! x_{42} \;, \nonumber \\
P( \nu_\mu\! \rightarrow \nu_\mu) & \!\!=\!\! & \!1 \!\!-\! \frac{c^2_1}{4} \sin^2 \!x _{21} \!\!-\! \frac{c^2_1}{2} \sin^2 \!x _{31} \!\!-\! \frac{(c_1 s_1)^2}{4} \sin^2 \!x _{41} \!\!-\! \frac{1}{2} \sin^2 \!x _{32} \!-\! \frac{s^2_1}{4}\sin^2 \!x _{42} \!\!-\! \frac{s^2_1}{2}\sin^2 \!x _{43}\nonumber \\ &  \!\!=\!\! & \!P( \nu_\tau \rightarrow \nu_\tau) \;, \nonumber \\ 
P(\nu_\mu \rightarrow \nu_e)\! & \!\!=\!\! & \!\frac{c^2_1}{2} \sin^2 \!x _{21} \!\!-\! \frac{(c_1 s_1)^2}{2} \sin^2 \!x _{41} \!\!+\! \frac{s^2_1}{2}\sin^2 \!x _{42}  \!=\! P(\nu_\tau \rightarrow \nu_e) \;, \nonumber \\
P(\nu_\mu \rightarrow \nu_\tau)\! &  \!\!=\!\! & \!- \frac{c^2_1}{4} \sin^2 \!x _{21} \!\!+\! \frac{c^2_1}{2} \sin^2 \!x _{31} \!\!-\! \frac{(c_1 s_1)^2}{4} \sin^2 \!x _{41} \!\!+\! \frac{1}{2} \sin^2 \!x _{32} \!\!-\! \frac{s^2_1}{4}\sin^2 \!x _{42} \!\!+\! \frac{s^2_1}{2}\sin^2 \!x _{43} \,, \nonumber \\
P(\nu_\mu\! \rightarrow \nu_{e_s})\! & \!\!=\!\! & \!(c_1 s_1)^2\sin^2 \!x _{41} = \!P(\nu_\tau \rightarrow \nu_{e_s}) \;, \nonumber \\
P(\nu_e \rightarrow \nu_{e_s})\! & \!\!=\!\! & \!2(c_1 s_1)^2\sin^2 \!x _{41} \;, \nonumber \\ 
P(\nu_{e_s\!}\! \rightarrow \nu_{e_s\!})\! & \!\!=\!\! & \!1 \!-\! 4(c_1 s_1)^2\sin^2 \!x _{41} \;.
\end{eqnarray}

\ni Hence, the probability summation rules

\vspace{-0.2cm}

\begin{eqnarray}
P(\nu_e \rightarrow \nu_e) \,+ P(\nu_e \rightarrow \nu_\mu) +P(\nu_e \rightarrow \nu_\tau) + P(\nu_e \rightarrow \nu_{e_s}) & = & 1 \;, \nonumber \\
P(\nu_\mu \rightarrow \nu_e) + P(\nu_\mu \rightarrow \nu_\mu) + P(\nu_\mu \rightarrow \nu_\tau) + P(\nu_\mu \rightarrow \nu_{e_s})\! & = & 1 \;, \nonumber \\
P(\nu_\tau \rightarrow \nu_e)\, + P(\nu_\tau \rightarrow \nu_\mu) + P(\nu_\tau \rightarrow \nu_\tau) + P(\nu_\tau \rightarrow \nu_{e_s\!})\! & = & 1 \;, \nonumber \\
P(\nu_{e_s}\! \rightarrow \nu_e)\! + P(\nu_{e_s}\! \rightarrow \nu_\mu) + P(\nu_{e_s}\! \rightarrow \nu_\tau)\! + P(\nu_{e_s} \rightarrow \nu_{e_s\!})\! & = & 1 
\end{eqnarray}

\ni hold, as it should be, for two sterile neutrinos $\nu_{\mu_s}$ and $\nu_{\tau_s}$ are completely decoupled due to $ s_2 = 0$ and $s_3 = 0$.

With the conjecture that $m^2_1 \simeq m^2_2$, implying $\Delta m^2_{41} \simeq \Delta m^2_{42}$ and $\Delta m^2_{31} \simeq \Delta m^2_{32}$, the first three Eqs. (23) can be rewritten approximately as

\vspace{-0.2cm}

\begin{eqnarray} 
P(\nu_e \rightarrow \nu_e) & \simeq &  1 -  c^2_1 \sin^2 x _{21} - (1+c_1^2) s_1^2 \sin^2 x _{42} \;, \nonumber \\
P( \nu_\mu \rightarrow \nu_\mu) & \simeq & 1 - \frac{1 + c^2_1}{2} \sin^2 x _{32} - \frac{c^2_1}{4} \sin^2 x _{21} - \frac{(1+c^2_1) s_1^2}{4} \sin^2 x _{42}  - \frac{s^2_1}{2}\sin^2 x _{43} \;, \nonumber \\ 
P(\nu_\mu \rightarrow \nu_e) & \simeq & \frac{c^2_1}{2} \sin^2 x _{21} + \frac{s^4_1}{2}\sin^2 x _{42} \;.
\end{eqnarray} 

\ni If $|\Delta m^2_{21}| \ll |\Delta m^2_{42}|$ and

\begin{equation} 
|\Delta m^2_{21}| = \Delta m^2_{\rm sol} \sim (10^{-5}\;\;{\rm or}\;\;10^{-7} \;\; {\rm or}\;\; 10^{-10})\;{\rm eV}^2  
\end{equation} 

\ni (for LMA or LOW or VAC solar solution, respectively) [1], then under the conditions of solar experiments the first Eq. (25) gives 

\begin{equation} 
P(\nu_e \rightarrow \nu_e)_{\rm sol} \simeq  1 -  c^2_1 \sin^2 (x _{21})_{\rm sol} - \frac{(1+c_1^2) s_1^2}{2}\;,\;\; c^2_1 = \sin^ 2 2\theta_{\rm sol} \stackrel{<}{\sim} 1 \;.  
\end{equation} 

\ni If $|\Delta m^2_{21}| \ll |\Delta m^2_{32}| \ll |\Delta m^2_{42}|\,,\, |\Delta m^2_{43}|$ and 

\begin{equation} 
|\Delta m^2_{32}| = \Delta m^2_{\rm atm} \sim 3.5\times10^{-3}\;{\rm eV}^2 \;, 
\end{equation} 

\ni then for atmospheric experiments the second Eq. (25) leads to

\begin{equation} 
P( \nu_\mu \rightarrow \nu_\mu)_{\rm atm}  \simeq  1 - \frac{1 + c^2_1}{2} \sin^2 (x _{32})_{\rm atm} - \frac{(3+c^2_1) s_1^2}{8}\;\;,\;\;\frac{1 + c^2_1}{2}  = \sin^2 2\theta_{\rm atm} \sim 1\;.  
\end{equation} 

\ni Eventually, if $|\Delta m^2_{21}| \ll |\Delta m^2_{42}|$ and 

\begin{equation} 
|\Delta m^2_{42}| = \Delta m^2_{\rm LSND} \sim 1 \;{\rm eV}^2 \;\;(e.g.) \;,
\end{equation} 

\ni then in the LSND experiment the third Eq. (25) implies

\begin{equation} 
P( \nu_\mu \rightarrow \nu_e)_{\rm LSND}  \simeq \frac{s^4_1}{2}\sin^2 (x _{42})_{\rm LSND}\;\;,\;\;\frac{s^4_1}{2} = \sin^2 2\theta_{\rm LSND} \sim 10^{-2}\;\;(e.g.).  
\end{equation} 

\ni Thus,

\begin{equation} 
s_1^2 \sim 0.141 \;,\; c_1^2 \sim 0.859 \;,\; \frac{1 + c^2_1}{2}  \sim 0.929 \;,\;   
\frac{(1+c_1^2) s_1^2}{2} \sim 0.131\;,\;\frac{(3+c_1^2) s_1^2}{8} \sim 0.0682\;,
\end{equation} 

\ni if the LNSD effect really exists and gets the amplitude $ s^4_1/2 \sim 10^{-2}$.

If the value $c^2_1 = \sin^ 2 2\theta_{\rm sol}  \sim 0.8$ or 0.9 or 0.7 (corresponding to LMA or LOW or VAC solar solution, respectively) [1] is {\it accepted}, then the amplitudes $\sin^ 2 2\theta_{\rm atm} = (1 + c^2_1)/2 \sim 0.9$ or 0.95 or 0.85 and $\sin^ 2 2\theta_{\rm LSND} = s^4_1/2 \sim (2$ or 0.5 or 4.5)$\times 10^{-2}$ are {\it predicted} for atmospheric and LSND experiments. 

Concluding, we can say that Eqs. (27), (29) and (31) are not inconsistent with solar, atmospheric and LSND experiments, respectively. Note that in Eqs. (27) and (29) there are constant terms that modify moderately the usual two--flavor formulae. The above equations, valid for $s_2 = 0$ and $s_3 = 0$, follow from the first three oscillation formulae (23), if {\it either}

\begin{equation} 
m^2_1 \simeq m^2_2 \ll m^2_3 \ll m^2_4 
\end{equation} 

\ni with

\begin{equation} 
m^2_3 \ll 1\;\;{\rm eV}^2\;\;,\;\; m^2_4 \sim 1\;\;{\rm eV}^2\;\;,\;\; \Delta m^2_{21} \sim (10^{-5} -10^{-10})\;{\rm eV}^2 \ll \Delta m^2_{32} \sim 10^{-3} \;{\rm eV}^2 
\end{equation} 

\ni {\it or}

\begin{equation} 
 m^2_4 \ll m^2_1 \simeq m^2_2 \simeq m^2_3 
\end{equation} 

\ni with

\begin{equation} 
m^2_3 \sim 1\;\;{\rm eV}^2\;\;,\;\; m^2_4 \ll 1\;\;{\rm eV}^2\;\;,\;\; \Delta m^2_{21} \sim (10^{-5} -10^{-10})\;{\rm eV}^2 \ll \Delta m^2_{32} \sim 10^{-3} \;{\rm eV}^2 \;.
\end{equation} 

\ni Here, we must have $m^2_2 \ll m^2_3 \ll m^2_4 \sim 1\;{\rm eV}^2$ or $m^2_4 \ll m^2_2 \simeq m^2_3 \sim 1\;{\rm eV}^2$, since $\Delta m^2_{32} \sim 10^{-3} \;{\rm eV}^2 \ll |\Delta m^2_{42}| \sim 1\;{\rm eV}^2 $. The second case $ m^2_4 \ll m^2_1 \simeq m^2_2 \simeq m^2_3 \sim 1 \;{\rm eV}^2 $, where the neutrino mass state $\nu_ 4$ induced by the sterile neutrino $\nu_{e_s}$ gets a vanishing mass, seems to be more natural than the first case $ m^2_1 \simeq m^2_2 \ll m_3^2 \ll m^2_4 \sim 1 \;{\rm eV}^2 $, where such a state gains a considerable amount of Majorana righthanded mass "for nothing". (This is so, unless one believes in the liberal maxim "whatever is not forbidden is allowed": the Majorana righthanded mass is not forbidden by the electroweak SU(2)$\times$U(1) symmetry, in contrast to Majorana lefthanded and Dirac masses requiring this symmetry to be broken, say, by a combined Higgs mechanism that becomes then the origin of these masses.) In the second case if, in addition, the masses $m_5 $ and $m_6$ connected with two decoupled sterile neutrinos are vanishing, {\it the specific Majorana lefthanded mass matrix $ M^{(L)}$ dominates over the whole neutrino mass matrix} $ M $. Such a Majorana lefthanded dominance may be the reason, why neutrino mixing appears to be qualitatively different from the more familiar down-quark mixing implied by the interplay of up- and down-quark Dirac mass matrices. Note also that, when looking for too close analogies between textures of neutrinos and charged leptons, one fails to describe adequately the observed neutrino oscillation effects (including the possible LSND effect) [6].

If $ s_2 = s_3 = 0$ and $m_4 = m_5 = m_6 = 0$, then writing $m_1 = m \,,\,m_2 = m + \delta m_{21}\,,\,m_3 = m + \delta m_{21}+ \delta m_{32}$ we can present the neutrino mass matrix (11) in the form $M = M^{(0)} + \delta M $, where

\begin{equation} 
M^{(0)} = m  \left( \begin{array}{cccccc} \frac{1}{2}(1+c^2_1) & \frac{1}{2\sqrt{2}_{~}} {s^2_1} & - \frac{1}{2 \sqrt{2}_{\,}} s^2_1 & -\frac{1}{\sqrt{2}_{~}} c_1 s_1 & 0 & 0 \\
\frac{1}{2\sqrt{2}_{~}} {s^2_1} & \frac{1}{4}(3 + c^2_1) & \frac{1}{4}s^2_1 & \frac{1}{2} c_1 s_1 & 0 & 0 \\
 -\frac{1}{2\sqrt{2}_{~}} {s^2_1} & \frac{1}{4}s^2_1 & \frac{1}{4}(3 + c^2_1) & -\frac{1}{2} c_1 s_1 & 0 & 0
\\ -\frac{1}{\sqrt{2}_{~}} {c_1 s_1} & \frac{1}{2} c_1 s_1 & -\frac{1}{2} c_1 s_1  & s^2_1 & 0 & 0 \\ 0 & 0 & 0 & 0 & 0 & 0  \\ 0  & 0 & 0  & 0  & 0 & 0 \end{array} \right) \;
\end{equation}

\ni is slightly modified by

\begin{equation} 
\,\delta M = \left( \begin{array}{cccccc} \frac{1}{2} \,\delta m_{21} & \frac{1}{2\sqrt{2}_{~}} \,\delta m_{21} & - \frac{1}{2 \sqrt{2}_{\,}} \,\delta m_{21} & 0 & 0 & 0 \\
\frac{1}{2\sqrt{2}_{~}} \,\delta m_{21} & \frac{1}{4}(3\,\delta m_{21} + 2\,\delta m_{32}) & \frac{1}{4} (\,\delta m_{21} + 2\,\delta m_{32}) & 0 & 0 & 0 \\
 -\frac{1}{2\sqrt{2}_{~}} \,\delta m_{21} & \frac{1}{4} (\,\delta m_{21} + 2\,\delta m_{32}) & \frac{1}{4}(3\,\delta m_{21} + 2\,\delta m_{32}) & 0 & 0 & 0 \\  0 & 0 & 0 & 0  & 0 & 0 \\ 0 & 0 & 0 & 0 & 0 & 0  \\ 0  & 0 & 0  & 0  & 0 & 0 \end{array} \right) \;.
\end{equation}

\ni In fact, $\delta m_{21} \sim m \,\delta m_{21}/{\rm eV} \simeq \Delta m^2_{21}/2{\rm eV} \sim 0.5 \times (10^{-5}$ or $10^{-7}$ or $10^{-10}$) eV and $\delta m_{32} \sim m \,\delta m_{32}/{\rm eV} \simeq \Delta m^2_{32}/2{\rm eV} \sim 1.5 \times 10^{-3}$ eV, while $ m \sim 1$ eV. In the formal limit of $ s_1 \rightarrow 0$, we obtain $M^{(0)}$ diagonal and degenerated in active and sterile neutrinos separately,

\begin{equation} 
M^{(0)} \rightarrow {\rm diag} ( m\,,\, m\,,\,m\,,\,0\,,\,0\,,\,0)\;,
\end{equation} 

\ni and so, from Eqs. (10) and (8) we infer that

\begin{equation} 
U^\dagger \,\delta M U \rightarrow {\rm diag} ( 0\,,\,\delta m_{21}\,,\, \delta m_{21} + \delta m_{32}\,,\,0\,,\,0\,,\,0)
\end{equation} 

\ni as $ U^\dagger M^{(0)} U \rightarrow$ diag$( m\,,\, m\,,\,m\,,\,0\,,\,0\,,\,0)$. Note from Eq. (8) that (with $s_2 = s_3 = 0$) in this limit {\it we get bimaximal mixing matrix $U$ in spite of the fact that in a good approximation the mass matrix $ M \simeq M^{(0)}$ is diagonal} (here, of course, the degeneracy of $\lim_{s_1 \rightarrow 0} M^{(0)}$ in active neutrinos works).

In the approximation used before to derive Eqs. (27), (29) and (31) there are true also the relations

\begin{eqnarray} 
P(\nu_e\, \rightarrow\, \nu_e)_{\rm sol} & \simeq & \!\!\!1 - \! P( \nu_e \rightarrow \nu_\mu)_{\rm sol}\! - \! P( \nu_e \rightarrow \nu_\tau)_{\rm sol}\! - \!(c_1 s_1)^2 \;, \; (c_1 s_1)^2 \sim 0.121 \;, \nonumber \\
P( \nu_\mu\! \rightarrow \nu_\mu)_{\rm atm} & \simeq & \!\!\!1 - P( \nu_\mu \rightarrow \nu_\tau)_{\rm atm} -  \frac{(1 + c^2_1)s^2_1}{4}\;\; , \;\; \frac{(1 + c^2_1) s^2_1}{4} \sim 0.0654 \;,  
\end{eqnarray} 

\ni as well as

\begin{equation} 
P(\nu_\mu \rightarrow \nu_e)_{\rm LSND} \simeq \frac{1}{2}\left( \frac{s_1}{c_1} \right)^2 P(\nu_\mu \rightarrow \nu_{e_s})_{\rm LSND} \;\;,\; \frac{1}{2}\left( \frac{s_1}{c_1}\right)^2 \sim 0.0824 \;\;.\;\;
\end{equation}

\ni The second relation (41) demonstrates a leading role of the appearance mode $\nu_\mu \rightarrow \nu_\tau $ in the disappearance process of atmospheric $\nu_\mu$'s, while the relation (42) indicates a direct interplay of the appearance modes $\nu_\mu \rightarrow \nu_e $ and $\nu_\mu \rightarrow \nu_{e_s} $. In the case of the first relation (41), both appearance modes $\nu_e \rightarrow \nu_\mu $  and $\nu_e \rightarrow \nu_\tau $ contribute equally to the disappearance process of solar $\nu_e$'s, and the role of the appearance mode $\nu_e \rightarrow \nu_{e_s}$ (responsible for the constant term) is also considerable.

Finally, for the Chooz experiment [7], where $(x_{ji})_{\rm Chooz} \simeq (x_{ji})_{\rm atm}$ for any $\Delta m^2_{ji}$, the first Eq. (25) predicts 

\begin{equation} 
P(\bar{\nu}_e \rightarrow\, \bar{\nu}_e)_{\rm Chooz}  \simeq  P( \bar{\nu}_e \rightarrow \bar{\nu}_e)_{\rm atm} \simeq 1 -  \frac{(1 + c^2_1)s^2_1}{2} \;, \;  \frac{(1 + c^2_1)s^2_1}{2} \sim 0.131 \;,
\end{equation} 

\ni if there is the LSND effect with the amplitude $s^4_1/2 \sim 10^{-2}$ as written in Eq. (31). Here, $ (1 + c^2_1)s^2_1\sin^2(x_{42})_{\rm Chooz} \simeq  (1 + c^2_1)s^2_1/2$. In terms of the usual two--flavor formula, the Chooz experiment excludes the disappearance process of reactor $ \bar{\nu}_e$'s for moving $\sin^2 2\theta_{\rm Chooz} \stackrel{>}{\sim} 0.1 $, when the range of moving $\Delta m^2_{\rm Chooz} \stackrel{>}{\sim} 3\times 10^{-3}\;{\rm eV}^2 $ is considered. In our case $\sin^2 2\theta_{\rm Chooz} \sim  (1 + c^2_1)s^2_1/2$ for $\sin^2 x_{\rm Chooz} \sim 1$. Thus, the Chooz effect for reactor $\bar{\nu}_e$'s should appear at the edge (if the LSND effect really exists). 

\vfill\eject

~~~~
\vspace{0.5cm}

{\centerline{\bf References}}

\vspace{0.5cm}

{\everypar={\hangindent=0.6truecm}
\parindent=0pt\frenchspacing

{\everypar={\hangindent=0.6truecm}
\parindent=0pt\frenchspacing

[1]~For a review {\it cf.} E. Kearns, Plenary talk at {\it ICHEP 2000} at Osaka; C.~Gonzales--Garcia, Talk at {\it ICHEP 2000} at Osaka.

\vspace{0.2cm}

[2]~G. Mills, Talk at {\it Neutrino 2000}; R.L.~Imlay, Talk at {\it ICHEP 2000} at Osaka; and references therein.

\vspace{0.2cm}

[3]~For a review {\it cf.} F. Feruglio, {\it Acta Phys. Pol.} {\bf B 31}, 1221 (2000); J.~Ellis, Summary of {\it Neutrino 2000}, hep-ph/0008334; and references therein.

\vspace{0.2cm}

[4]~W. Kr\'{o}likowski, hep--ph/0102016.

\vspace{0.2cm}

[5]~W. Kr\'{o}likowski, hep--ph/0007255.

\vspace{0.2cm}

[6]~W. Kr\'{o}likowski, {\it Nuovo Cim.} {\bf A 111}, 1257 (1998); {\it Acta Phys. Pol.} {\bf B 31}, 1759 (2000).

\vspace{0.2cm}

[7]~M. Appolonio {\it et al.}, {\it Phys. Lett.} {\bf B 420}, 397 (1998); {\bf B 466}, 415 (1999).

\vfill\eject

\end{document}